\documentclass[12pt,a4paper,final]{iopart}
\usepackage{iopams}

\expandafter\let\csname equation*\endcsname\relax
\expandafter\let\csname endequation*\endcsname\relax

\usepackage{amsmath}
\usepackage{amsfonts}
\usepackage{amsmath}
\usepackage{amssymb}
\usepackage{amsthm}
\usepackage[utf8]{inputenc}
\usepackage{graphicx}
\usepackage[table]{xcolor}
\usepackage{bm}
\usepackage{booktabs}
\usepackage{cite}

\usepackage[breaklinks=true,colorlinks=true,linkcolor=blue,urlcolor=blue,citecolor=blue]{hyperref}

\newcommand{\defeq}{\mathrel{\mathop:}=}
\newcommand{\Var}[2]{\text{Var}\left(#1; #2\right)}

\newcommand{\U}{\mathcal{U}}
\newcommand{\bF}{\beta_F}
\newcommand{\bFs}{b_F}
\newcommand{\bFe}{\mathcal{B}_F}
\newcommand{\bO}{\beta_\Omega}
\newcommand{\bOs}{b_\Omega}
\newcommand{\bOe}{\mathcal{B}_\Omega}
\newcommand{\dF}{\big<{\bFs}'\big>_S}

\newcommand{\DF}{\big<{\bF}'\big>_S}
\newcommand{\Es}{h}
\newcommand{\Ee}{g}

\DeclareMathOperator*{\sgn}{sgn}

\begin{document}

\title{A classification of nonequilibrium steady states based on temperature correlations}

\author[cor1]{Sergio Davis$^{1,2}$}
\address{$^1$Research Center on the Intersection in Plasma Physics, Matter and Complexity, P$^2$mc, Comisión Chilena de Energía Nuclear, Casilla 188-D, Santiago, Chile}
\address{$^2$Departamento de F\'isica, Facultad de Ciencias Exactas, Universidad Andres Bello. Sazi\'e 2212, piso 7, 8370136, Santiago, Chile.}
\ead{sergio.davis@cchen.cl}

\begin{abstract}
Although generalized ensembles have now been in use in statistical mechanics for decades, including frameworks such as Tsallis' nonextensive statistics and superstatistics,
a classification of these generalized ensembles outlining the boundaries of validity of different families of models, is still lacking. In this work, such a classification
is proposed in terms of \emph{supercanonical} and \emph{subcanonical} ensembles, according to a newly defined parameter, the inverse temperature covariance parameter $\U$.
This parameter is non-negative in superstatistics (and is equal to the variance of the inverse temperature) but can be negative for other families of statistical ensembles,
adquiring then a broader meaning. It is shown that $\U$ is equal for every region of a composite system in a steady state, and examples are given of supercanonical and
subcanonical states.
\end{abstract}

\section{Introduction}

Ensembles beyond the canonical ensemble of equilibrium thermodynamics have already been used for decades. On the one hand, they have been successful in describing
nonequilibrium systems in steady states such as plasmas~\cite{Lima2000,Ourabah2015,Davis2019b,Ourabah2020b}, self-gravitating systems~\cite{Lima2002,Iguchi2005,Nardini2012}
and other complex systems~\cite{Latora2001,Silva2006b,Atenas2021}. On the other hand, they have extended existing computer simulation methods~\cite{Berg2002}. Among these
generalized ensembles, the $q$-canonical ensemble, commonly associated with Tsallis' nonextensive statistics~\cite{Tsallis2009c}, is widely used in the description of a variety
of complex systems that follow power laws~\cite{Tsallis2009}. Instead of the traditional canonical ensemble
\begin{equation}
\label{eq:canon}
P(\bm x|\beta) = \frac{\exp(-\beta \mathcal{H}(\bm x))}{Z(\beta)},
\end{equation}
that describes thermal equilibrium for a system with Hamiltonian $\mathcal{H}$ at inverse temperature $\beta = 1/(k_B T)$, the $q$-canonical ensemble
considers a microstate distribution of the form
\begin{equation}
\label{eq:qcanon}
P(\bm x|q, \beta) = \frac{\exp(-\beta \mathcal{H}(\bm x); q)}{Z_q(\beta)},
\end{equation}
having an additional parameter $q$, where
\begin{equation}
\exp(x; q) \defeq \big[1+(1-q)x\big]_+^{\frac{1}{1-q}}
\end{equation}
is the $q$-exponential function. Because $\exp(x; 1)=\exp(x)$, the limit $q \rightarrow 1$ of \eqref{eq:qcanon} recovers \eqref{eq:canon}, so the canonical ensemble is
contained as a particular case.

Despite the success of these generalized ensembles, there are still fundamental questions about their range of validity. For instance, it has been clearly established~\cite{Lutsko2011}
that states with $q > 1$ are qualitatively different than those with $q < 1$, and have been classified as subadditive and superadditive, respectively~\cite{Bagci2016}. Moreover,
only states with $q \geq 1$ can be described using superstatistics~\cite{Beck2003,Beck2004}, an elegant framework which assumes that temperature is a random variable with distribution
$P(\beta|S)$, such that the joint distribution of $\beta$ and the microstate $\bm x$ is given by
\begin{equation}
\label{eq:super_joint}
P(\bm x, \beta|S) = P(\bm x|\beta, S)P(\beta|S) = \left[\frac{\exp(-\beta\mathcal{H}(\bm x))}{Z(\beta)}\right] P(\beta|S).
\end{equation}

Beyond the studies regarding the $q$-canonical ensemble, however, no attempt has been made to classify the nonequilibrium steady states in general, or to search for any general
explanation of the differences between the two separate regimes of $q$-canonical states. In this work we propose such a classification of nonequilibrium steady states into two
large categories, namely supercanonical and subcanonical states, using a statistical measure of the covariance between temperature fluctuations.

This paper is organized as follows. In Section \ref{sec:steady} we first review two definitions of temperature for generalized ensembles, then in Sections \ref{sec:tempcov} and
\ref{sec:micro} we introduce the inverse temperature covariance $\U$, in terms of which we will express our main results, and present the proposed classification of nonequilibrium
steady states. Section \ref{sec:examples} presents some examples of nonequilibrium steady states in each category, and we provide some concluding remarks in Section
\ref{sec:concluding}.

\section{Steady states, temperatures and superstatistics}
\label{sec:steady}

We will define a generalized ensemble $S$ as any non-equilibrium steady state with microstate probabilities of the form
\begin{equation}
\label{eq:rho}
P(\bm x|S) = \rho(\mathcal{H}(\bm x); S),
\end{equation}
where the function $\rho$ is the ensemble function associated to $S$. If an arbitrary value of $\mathcal{H}$ is denoted by $E$, the energy distribution in this ensemble
is given by
\begin{equation}
\label{eq:energydist}
P(\mathcal{H} = E|S) = \rho(E; S)\Omega(E)
\end{equation}
with
\begin{equation}
\Omega(E) \defeq \int d\bm{x}\delta(E-\mathcal{H}(\bm x))
\end{equation}
the density of states. A particular class among these steady states is described by superstatistics, as defined by \eqref{eq:super_joint}. By integrating over $\beta$ we obtain
the microstate distribution
\begin{equation}
\label{eq:super_marg}
P(\bm x|S) = \int_0^\infty d\beta P(\bm x, \beta|S) = \int_0^\infty d\beta P(\beta|S)\left[\frac{\exp(-\beta\mathcal{H}(\bm x))}{Z(\beta)}\right],
\end{equation}
and therefore the superstatistical ensemble function is
\begin{equation}
\rho(E; S) = \int_0^\infty d\beta P(\beta|S)\left[\frac{\exp(-\beta E)}{Z(\beta)}\right].
\end{equation}

From \eqref{eq:super_marg} it is clear that the canonical ensemble at $\beta_0$ is recovered from superstatistics when $\beta_0$ is the only possible value of temperature,
that is, when
\begin{equation}
P(\beta|\beta_0) = \delta(\beta-\beta_0).
\end{equation}
Every steady state defined by \eqref{eq:rho} has a \emph{fundamental inverse temperature} function $\bF(E; S)$, given by
\begin{equation}
\label{eq:betaf}
\bF(E; S) \defeq -\frac{\partial}{\partial E}\ln \rho(E; S)
\end{equation}
and such that the canonical ensemble is the only ensemble where $\bF$ is a constant function, $\bF(E; \beta_0) = \beta_0$. Another definition of temperature, this time
being intrinsic to the Hamiltonian, i.e. independent of the ensemble function, is provided by the \emph{microcanonical inverse temperature},
\begin{equation}
\bO(E) \defeq \frac{\partial}{\partial E}\ln \Omega(E),
\end{equation}
also known in the community of computer simulation as the \emph{statistical temperature}~\cite{Kim2006b}. This temperature corresponds to the thermodynamic definition
\begin{equation}
\frac{1}{T} \defeq \frac{\partial S}{\partial E},
\end{equation}
with $S(E) \defeq k_B\ln \Omega(E)$ the Boltzmann entropy. Note that the most probable energy, denoted by $E^*$, is a solution of the extremum equation
\begin{equation}
\label{eq:extremum}
0 = \frac{\partial}{\partial E}\ln P(\mathcal{H} = E|S)\Big|_{E=E^*} = \beta_F(E^*) - \beta_\Omega(E^*),
\end{equation}
therefore it is such that it equalizes the fundamental and microcanonical inverse temperatures. The specific heat $C_E$ at constant energy is connected to this
microcanonical temperature through the relation
\begin{equation}
\label{eq:specheat}
C_E \defeq \left(\frac{\partial T(E)}{\partial E}\right)^{-1} = -\frac{\bO(E)^2}{{\bO}'(E)},
\end{equation}
and is in general a function of $E$. From this definition is clear that the sign of $C_E$ is always the opposite of the sign of the derivative ${\bO}'(E)$.

By assuming that the energy distribution in \eqref{eq:energydist} has zero probability at its boundaries, and invoking the conjugate variables
theorem (CVT)~\cite{Davis2012,Davis2016c}, we obtain the identity
\begin{equation}
\label{eq:cvt}
\left<\frac{\partial \omega}{\partial E}\right>_S = -\left<\omega\frac{\partial}{\partial E}\ln P(E|S)\right>_S = \Big<\omega\big(\bF-\bO)\Big>_S,
\end{equation}
where $\omega=\omega(E)$ is an arbitrary, differentiable function of the energy. Taking $\omega(E) = 1$ we readily see that
\begin{equation}
\big<\bF\big>_S = \big<\bO\big>_S,
\end{equation}
and we will refer to this mean inverse temperature as the inverse temperature $\beta_S$ of the ensemble,
\begin{equation}
\beta_S \defeq \big<\bO\big>_S = \big<\bF\big>_S.
\end{equation}

\noindent
In the case of superstatistics, it follows directly from the definition of $\beta_F$ in \eqref{eq:betaf} that
\begin{equation}
\label{eq:betaFE}
\beta_F(E; S) = -\frac{\partial}{\partial E}\ln \rho(E; S) = \int_0^\infty d\beta \left[\frac{P(\beta|S)\exp(-\beta E)}{\rho(E; S)Z(\beta)}\right]\beta = \big<\beta\big>_{E,S},
\end{equation}
where we have recognized the conditional distribution
\begin{equation}
P(\beta|E, S) = \frac{P(E, \beta|S)}{P(E|S)} = \frac{P(\beta|S)\exp(-\beta E)}{\rho(E; S)Z(\beta)}.
\end{equation}

\noindent
By taking expectation of \eqref{eq:betaFE} in the form
\begin{equation}
\big<\beta\big>_{E,S} = \beta_F(E),
\end{equation}
using $P(\mathcal{H} = E|S)$, we can also assert that $\big<\beta\big>_S = \beta_S$, where the expectation of the left-hand side is the mean superstatistical inverse temperature.

\section{The inverse temperature covariance $\U$}
\label{sec:tempcov}

In previous works~\cite{Davis2020} it was shown that the variance of the superstatistical inverse temperature $\beta$ can, in principle, be measured through the equality
\begin{equation}
\label{eq:varsuper}
\Var{\beta}{S} = \Var{\bO}{S} + \big<{\bO}'\big>_S.
\end{equation}
where $\Var{A}{S}$ denotes the variance of the quantity $A$ in the steady state $S$, defined as
\begin{equation}
\Var{A}{S} \defeq \big<(\delta A)^2\big>_S = \big<A^2\big>_S - \big<A\big>_S^2.
\end{equation}

\noindent
The quantity in the right-hand side of \eqref{eq:varsuper},
\begin{equation}
\label{eq:udef}
\U \defeq \Var{\bO}{S} + \big<{\bO}'\big>_S,
\end{equation}
only involves the statistics of the microcanonical inverse temperature and, according to \eqref{eq:varsuper}, is non-negative in the case of superstatistics.

\newpage

Because the variance of $\beta$ is equal to zero only in the canonical ensemble, that case corresponds to $\U = 0$. However, the condition $\U \geq 0$ does not always hold
for more general steady states, as $\U$ has been shown to take negative values in certain states outside superstatistics~\cite{Davis2022}.

The fact that $\U$ can be either positive or negative, and furthermore, that $\U = 0$ for the canonical ensemble, suggests a classification of steady states into
\emph{supercanonical} states, where $\U > 0$, and \emph{subcanonical} states, with $\U < 0$. Although in the subcanonical case $\U$ can no longer be associated with a variance,
it is always equal to the covariance between $\bF$ and $\bO$,
\begin{equation}
\label{eq:udef2}
\U = \big<\delta\bF\delta\bO\big>_S
\end{equation}
which can be negative. This follows directly by choosing $\omega(E) = \bO(E)$ in \eqref{eq:cvt}. By instead choosing $\omega(E) = \bF(E)$ we obtain an alternative expression,
this time in terms of the fundamental inverse temperature, namely
\begin{equation}
\label{eq:ubetaf}
\U = \Var{\bF}{S} - \big<{\bF}'\big>_S,
\end{equation}
by which we can immediately see that a constant $\beta_F$ function, having zero variance and zero derivative, implies $\U = 0$. By using $\omega(E) = \bF(E)-\bO(E)$
in \eqref{eq:cvt}, we obtain
\begin{equation}
\big<{\bF}'\big>_S - \big<{\bO}'\big>_S = \big<(\bF-\bO)^2\big>_S \geq 0
\end{equation}
hence we have the inequality
\begin{equation}
\label{eq:ineq}
\big<{\bF}'\big>_S \geq \big<{\bO}'\big>_S.
\end{equation}

\noindent
From this inequality and using \eqref{eq:udef} and \eqref{eq:ubetaf} it follows that
\begin{equation}
\Var{\bF}{S} > \Var{\bO}{S} \qquad\text{for}\; \big<{\bO}'\big>_S > 0,
\end{equation}
while on the other hand,
\begin{equation}
\Var{\bO}{S} > \Var{\bF}{S} \qquad\text{for}\; \big<{\bF}'\big>_S < 0.
\end{equation}

In both these cases where the ordering of the variances is well established we see, by replacing in \eqref{eq:udef} or \eqref{eq:ubetaf} respectively, that $\U > 0$, which
means there are two clearly disjoint regimes within the supercanonical states, that we will refer to as supercanonical-A when the variance of $\beta_F$ is higher, and
supercanonical-B otherwise. No such clear separation ocurrs for the subcanonical ($\U < 0$) states: they only require that $\big<{\bF}'\big>_S > 0$ and $\big<{\bO}'\big>_S < 0$,
and are therefore always consistent with \eqref{eq:ineq}.

Table~\ref{tbl:classif} shows the proposed classification of steady states in this work. All of superstatistics, including the canonical ensemble, is contained in the
supercanonical B class, so we see that there are three entire categories outside superstatistics.

\begin{table}
\begin{center}
\begin{tabular}{l|c|c|c|c|}
\cline{2-5} & Sign of $\U$ & Signature & Ordering of variances & Sign of $C_E$ \\
\cline{1-5}
\multicolumn{1}{|l|}{Supercanonical A} & $\U  >   0$ & $(+,+)$ & $\Var{\bF}{S} > \Var{\bO}{S}$  & $C_E < 0$ \\ \cline{1-5}
\multicolumn{1}{|l|}{Supercanonical B} & $\U \geq 0$ & $(-,-)$ & $\Var{\bO}{S} > \Var{\bF}{S}$  & $C_E > 0$ \\ \cline{1-5}
\multicolumn{1}{|l|}{Subcanonical A}   & $\U \leq 0$ & $(-,+)$ & $\Var{\bF}{S} > \Var{\bO}{S}$ & $C_E > 0$  \\ \cline{1-5}
\multicolumn{1}{|l|}{Subcanonical B}   & $\U \leq 0$ & $(-,+)$ & $\Var{\bO}{S} > \Var{\bF}{S}$  & $C_E > 0$ \\ \cline{1-5}
\end{tabular}
\end{center}
\caption{Defining features of the classification of steady states proposed in this work, consisting of supercanonical and subcanonical systems. Superstatistical
states, including the canonical ensemble, belong to the supercanonical-B class. The \emph{signature} of a state is the sign of $\big<{\bO}'\big>_S$ followed by
the sign of $\big<{\bF}'\big>_S$, where the signature $(+, -)$ is forbidden by the inequality in \eqref{eq:ineq}.}
\label{tbl:classif}
\end{table}

\newpage
\subsection{Subsystem and environment}

Let us consider a composite system in a nonequilibrium steady state. Its Hamiltonian $\mathcal{H}$ is given by the sum of two terms, the energy $H(\bm x)$ of the
subsystem $\bm x$ and the energy $G(\bm y)$ of the environment $\bm y$, that is,
\begin{equation}
\mathcal{H}(\bm x, \bm y) = H(\bm x) + G(\bm y),
\end{equation}
so that the distribution of microstates is
\begin{equation}
P(\bm x, \bm y|S) = \rho\big(H(\bm x)+G(\bm y); S\big).
\end{equation}

Here $\bm{x}$ and $\bm{y}$ denote the entire set of degrees of freedom, both kinetic and configurational. As the notation becomes intricate for all the quantities in subsystem
and environment, a summary is presented in Table~\ref{tbl:notation}.

The joint distribution of the energy $h$ of the subsystem and the energy $g$ of the environment is given by
\begin{equation}
\begin{split}
P(h, g|S) & = \Big<\delta(h-H(\bm x))\delta(g-G(\bm y))\Big>_S \\
          & = \int d\bm{x}d\bm{y}\:\rho(H(\bm x)+G(\bm y); S)\:\delta(h-H(\bm x))\delta(g-G(\bm y)) \\
          & = \rho(h+g; S)\Omega_H(h)\Omega_G(h),
\end{split}
\end{equation}
where $\Omega_H$ and $\Omega_G$ are the densities of states of subsystem and environment, respectively. We can see that these energies are in general correlated
unless $\rho(h+g)$ is separable, which is the case for the canonical ensemble. The general form of the fluctuation-dissipation theorem for the derivative of
expectations~\cite{Davis2016c} gives, for an arbitrary function $\omega(h, g)$, the identity
\begin{equation}
\label{eq:fdt}
\frac{\partial}{\partial h}\big<\omega\big>_{h,S} = \left<\frac{\partial \omega}{\partial h}\right>_{h,S} + \left<\omega\frac{\partial}{\partial h}\ln P(g|h, S)\right>_{h,S},
\end{equation}
which upon replacing
\begin{equation}
\label{eq:PghS}
P(g|h, S) = \frac{P(h, g|S)}{P(h|S)} = \frac{\rho(h+g; S)\Omega_G(g)}{\rho_h(h; S)}
\end{equation}
gives, for the logarithmic derivative,
\begin{equation}
\frac{\partial}{\partial h}\ln P(g|h, S) = \bFs(h)-\bF(h+g; S),
\end{equation}
and reduces \eqref{eq:fdt} to
\begin{equation}
\label{eq:fdt2}
\frac{\partial}{\partial h}\big<\omega\big>_{h,S} = \left<\frac{\partial \omega}{\partial h}\right>_{h,S} + \big<\omega\big(\bFs-\bF\big)\big>_{h,S}.
\end{equation}

\noindent
In a similar way, applying the CVT to $P(g|h, S)$ and using the logarithmic derivative
\begin{equation}
\frac{\partial}{\partial g}\ln P(g|h, S) = \bOe(g) - \bF(h+g; S),
\end{equation}
we obtain
\begin{equation}
\label{eq:cvt2}
\left<\frac{\partial \omega}{\partial g}\right>_{h, S} = \big<\omega\big(\bF - \bOe\big)\big>_{h, S}.
\end{equation}

\noindent
Setting $\omega(h, g) = 1$ in \eqref{eq:fdt2} we readily obtain
\begin{equation}
\big<\bFs-\bF\big>_{h,S} = 0,
\end{equation}
and because $\bFs$ is only a function of $h$, we can write
\begin{equation}
\label{eq:betaf_proj}
\big<\bF\big>_{h,S} = \bFs(h).
\end{equation}

This rule expresses the fundamental inverse temperature of the subsystem as a conditional mean of the fundamental inverse temperature of the entire system.
Taking expectation over $h$ in the state $S$ we see that
\begin{equation}
\label{eq:betas_eq}
\big<\bFs\big>_S = \big<\bF\big>_S = \beta_S.
\end{equation}

In other words, the inverse temperature of the composite system must be equal to the inverse temperature of the subsystem. Repeating the argument but using
$g$ instead of $h$, we see that
\begin{equation}
\big<\bFs\big>_S = \big<\bFe\big>_S = \beta_S,
\end{equation}
and this is an extension of the defining property of temperature as the quantity that equalizes in equilibrium.  Additionally, using \eqref{eq:cvt2} with $\omega(h, g) = 1$
gives the important result
\begin{equation}
\label{eq:betao_proj}
\big<\bOe\big>_{h, S} = \big<\bF\big>_{h,S} = \bFs(h),
\end{equation}
where the last equality is due to \eqref{eq:betaf_proj}. This new relation in \eqref{eq:betao_proj} is a generalization of
\begin{equation}
\bFs(h) = \bOe(E-h)
\end{equation}
for the case where $h+g=E$ is fixed, as mentioned in Ref.~\cite{Davis2022}. It implies that the fundamental (inverse) temperature of a subsystem is measurable
if we have access to the conditional energy distribution $P(g|h, S)$.

Now we have all the elements to evaluate the inverse temperature covariance $\U_h$ for the subsystem. Using the choice $\omega(h, g) = \bF(h+g; S)$ in \eqref{eq:fdt} gives
\begin{equation}
\big<{\bF}'\big>_{h,S} = \big<{\bF}^2\big>_{h,S} - \bFs(h)^2 + {\bFs}'(h),
\end{equation}
and taking expectation under $S$ it follows that
\begin{equation}
\DF = \big<{\bF}^2\big>_S - \big<(\bFs)^2\big>_S + \dF.
\end{equation}

\noindent
Substracting $\beta_S$ from both sides and using \eqref{eq:betas_eq}, we finally obtain
\begin{equation}
\Var{\bF}{S} - \DF = \Var{\bFs}{S} - \dF,
\end{equation}
that is, $\U = \U_h$. Repeating the same analysis for $g$ instead of $h$, it must also hold that $\U = \U_g$, so we have
\begin{equation}
\label{eq:equalU}
\U = \U_h = \U_g.
\end{equation}

\begin{table}
\begin{center}
\begin{tabular}{|c|c|c|}
\hline
Property & Domain & Notation \\
\hline
Hamiltonian & Entire system & $\mathcal{H}(\bm x, \bm y)$ \\
Hamiltonian & Subsystem     & $H(\bm x)$ \\
Hamiltonian & Environment   & $G(\bm y)$ \\
\hline
Value of energy & Entire system & $E$ \\
Value of energy & Subsystem     & $h$ \\
Value of energy & Environment   & $g$ \\
\hline
Fundamental $\beta$ & Entire system & $\bF(E)$ \\
Fundamental $\beta$ & Subsystem     & $\bFs(h)$ \\
Fundamental $\beta$ & Environment   & $\bFe(g)$ \\
\hline
Microcanonical $\beta$ & Entire system & $\bO(E)$ \\
Microcanonical $\beta$ & Subsystem     & $\bOs(h)$ \\
Microcanonical $\beta$ & Environment   & $\bOe(g)$ \\
\hline
Covariance of $\beta$ & Entire system & $\U$ \\
Covariance of $\beta$ & Subsystem     & $\U_h$ \\
Covariance of $\beta$ & Environment   & $\U_g$ \\
\hline
\end{tabular}
\end{center}
\caption{Notation used in this work when describing a composite system and its parts. The quantity $\U$ is the inverse temperature
covariance defined by \eqref{eq:udef2}.}
\label{tbl:notation}
\end{table}

This is a new and important result for nonequilibrium steady states, which tells us that in a steady state, the inverse temperature covariance $\U$ is always
the same for the composite system, the subsystem and the environment. This result may help explain the findings by Nauenberg~\cite{Nauenberg2003}, that subsystems
with different entropic index $q$ cannot reach equilibrium, or more generally, reach a steady state, provided that different values of $q$ produce different
values of $\U$, as in the example presented in subsection \ref{sec:tsallis}. From \eqref{eq:equalU} we can readily obtain the classical result by Ray~\cite{Ray1991b}
that small systems in the microcanonical ensemble have non-Maxwellian momentum distributions: if $\U \neq 0$ for the configurational degrees of freedom, the kinetic
degrees of freedom cannot follow a canonical distribution with $\U = 0$.

In the case of superstatistics it is clear why \eqref{eq:equalU} is true: for the composite system we have
\begin{equation}
\rho(h+g; S) = \int_0^\infty d\beta P(\beta|S)\frac{\exp(-\beta(h+g))}{Z(\beta)},
\end{equation}
so the ensemble function of the subsystem is obtained by marginalization as
\begin{equation}
\begin{split}
\rho_h(h; S) & = \int dg\:\Omega_G(g)\rho(h+g; S) \\
             & = \int_0^\infty d\beta P(\beta|S)\frac{\exp(-\beta h)}{Z(\beta)}\int dg\Omega_G(g)\exp(-\beta g) \\
             & = \int_0^\infty d\beta P(\beta|S)\frac{\exp(-\beta h)}{Z_h(\beta)},
\end{split}
\end{equation}
where we have used $Z(\beta)=Z_h(\beta)Z_g(\beta)$ and \[Z_g(\beta) = \int dg\Omega_G(g)\exp(-\beta g).\]

In other words, the subsystem must be described by the same $P(\beta|S)$ as the composite system and the environment. Because for superstatistics $\U$ coincides with
the variance of $\beta$, and is therefore determined by $P(\beta|S)$, it follows that \eqref{eq:equalU} must hold.

Furthermore, one can also prove that $\U$ has yet another definition, this time connecting subsystem and environment, namely
\begin{equation}
\label{eq:Uinternal}
\U = \big<\delta\bOs\delta\bOe\big>_S,
\end{equation}
that is, $\U$ is the covariance between the microcanonical inverse temperatures of subsystem and environment. In order to see why this is true, we write the
covariance in \eqref{eq:Uinternal} in terms of the joint distribution $P(h, g|S)$,
\begin{equation}
\begin{split}
\big<\delta \bOs \delta\bOe\big>_S & = \int dh dg P(h, g|S)\bOs(h)\bOe(g) - (\beta_S)^2 \\
& = \int dh P(h|S) \bOs(h)\left[\int dg P(g|h, S)\bOe(g)\right] - (\beta_S)^2 \\
& = \int dh P(h|S) \bOs(h) \bFs(h) - (\beta_S)^2 \\
& = \big<\delta\bOs \delta\bFs\big>_S = \U_h = \U,
\end{split}
\end{equation}
where we have used \eqref{eq:betao_proj} combined with \eqref{eq:betaf_proj} to replace the integral in square brackets. Also,
because of the law of total variance
\begin{equation}
\Var{X}{I} = \big<\Var{X}{\bm{Y},I}\big>_I + \Var{\big<X\big>_{\bm{Y},I}}{I},
\end{equation}
valid for any random scalar $X$ and random vector $\bm Y$, we have that, for any set of parameters $\bm{\theta}$,
\begin{equation}
\begin{split}
\U - \big<{\bO}'\big>_S & = \Var{\bO}{S} \\
                        & = \big<\Var{\bO}{\bm{\theta},S}\big>_S + \Var{\beta(\bm \theta)}{S} \\
                        & = \Big<\U_{\bm{\theta}} - \big<{\bO}'\big>_{\bm{\theta},S}\Big>_S + \Var{\beta(\bm \theta)}{S} \\
                        & = \big<\U_{\bm{\theta}}\big>_S - \big<{\bO}'\big>_S + \Var{\beta(\bm \theta)}{S}
\end{split}
\end{equation}
where we have defined $\beta(\bm \theta) \defeq \big<\bO\big>_{\bm{\theta},S}$ and used the property
\begin{equation}
\Big<\big<{\bO}'\big>_{\bm{\theta},S}\Big>_S = \big<{\bO}'\big>_S,
\end{equation}
hence we have the inequality
\begin{equation}
\U \geq \big<\U_{\bm{\theta}}\big>_S.
\end{equation}

Not only this indicates that $\U$ cannot decrease when lifting constraints, it gives a procedure to compute a lower bound for $\U$ in an arbitrary steady state
if we can compute $\U_{\bm \theta}$ for some fixed set of parameters $\bm{\theta}$, for instance, in microcanonical steady states. An important result for
the value of $\U$ in microcanonical, short-range systems is given in the next section.

\section{Regions of a large, microcanonical short-range system}
\label{sec:micro}

Consider an homogeneous isolated system where $\mathcal{H}(\bm x, \bm y) = H(\bm{x})+G(\bm{y}) = E$ with $E$ its fixed total energy. Here $\bm{x}$ and $\bm{y}$ form a
partition of the system into a subsystem and an environment, respectively. The joint distribution of $\bm{x}$ and $\bm{y}$ is microcanonical, so we have
\begin{equation}
P(\bm{x}, \bm{y}|E) = \frac{1}{\Omega(E)}\delta(H(\bm{x})+G(\bm{y})-E).
\end{equation}

In order to prove a simple but important result, we will make some assumptions. First, the microcanonical inverse temperatures associated to each region
$\bm{x}$ and $\bm{y}$ will be given by the functions
\begin{subequations}
\begin{align}
\bOs(\Es) & = f_1(\Es), \\
\bOe(\Ee) & = f_2(\Ee),
\end{align}
\end{subequations}
which are both strictly monotonic with $E$ and follow the same trend, i.e.,
\begin{equation}
\sgn(f_1') = \sgn(f_2').
\end{equation}

Second, the energy fluctuations of the subsystem are sufficiently small that we can consider $\Es = h^* + \delta \Es$ with small $\delta \Es$. Under these
conditions we will show that the microcanonical system must be subcanonical, that is,
\begin{equation}
\label{eq:Uneg}
\U = \big<\delta \bOs\delta \bOe\big>_{E} \leq 0,
\end{equation}
therefore $\U_h \leq 0$ and $\U_g \leq 0$, according to \eqref{eq:equalU}. This follows directly from \eqref{eq:Uinternal}, which we can write in the form
\begin{equation}
\label{eq:Unegproof}
\U = \Big<\big[\bOs-\beta_S\big]\big[\bOe-\beta_S\big]\Big>_{E}
\end{equation}
where $\beta_S = \big<\bOs\big>_{E} = \big<\bOe\big>_{E}$. Considering $\delta \Es$ to be small, this expectation can be approximated as
\begin{equation}
\beta_S = \big<f_1(\Es)\big>_{E} \approx \big<f_1(h^*) + \delta \Es f_1'(h^*)\big>_{E} = \beta_0 + \big<\delta \Es\big>_{E}f_1'(h^*) = \beta_0
\end{equation}
with $\beta_0 \defeq f_1(h^*)$, but in the same way
\begin{equation}
\beta_S = \big<f_2(\Ee)\big>_{E} \approx \big<f_2(E-h^*) + \delta \Ee f_2'(E-h^*)\big>_{E} = f_2(E-h^*)
\end{equation}
thus $\beta_0 = f_2(E-h^*)$. The fluctuations of temperature in subsystem and environment are given by
\begin{subequations}
\begin{align}
\delta \bOs = f_1(\Es) - \beta_0 & \approx \delta \Es f_1'(h^*), \\
\delta \bOe = f_2(E-\Es) - \beta_0 & \approx -\delta \Es f_2'(E-h^*).
\end{align}
\end{subequations}
where we have used $\delta \Es + \delta \Ee = \delta E = 0$. It is clear that these fluctuations always have opposite signs, therefore
\begin{equation}
\delta \bOs\delta \bOe = -(\delta \Es)^2 f_1'(h^*)f_2'(E-h^*) \leq 0
\end{equation}
for every value of $\Es$, and \eqref{eq:Uneg} follows by taking expectation. From Table \ref{tbl:classif} it also follows that $C_E > 0$ for the composite system
and its parts, so not only are $f_1$ and $f_2$ strictly monotonic with $E$, it must be true that $f_1' < 0$ and $f_2' < 0$.

\section{Examples}
\label{sec:examples}

\subsection{The $q$-canonical ensemble}
\label{sec:tsallis}

While we already know that $q \geq 1$ is described by the $\chi^2$-superstatistics~\cite{Beck2004}, the behavior of the $q$-canonical ensemble for $q \leq 1$ remains
to be explored. We can show that states with $q \leq 1$ can be subcanonical, that is, can have $\U < 0$.

\noindent
We start with the fundamental inverse temperature corresponding to \eqref{eq:qcanon}, given by~\cite{Davis2019}
\begin{equation}
\beta_F(E; q, \beta_0) = \frac{\beta_0}{1+(q-1)\beta_0 E},
\end{equation}
and compute its derivative,
\begin{equation}
{\beta_F}'(E; q, \beta_0) = (1-q)\beta_F(E; q, \beta_0)^2.
\end{equation}

We clearly see that this derivative is negative for $q > 1$, consistent with superstatistics, but positive for $q < 1$. The inverse temperature covariance is
\begin{equation}
\U = \big<(\delta \beta_F)^2\big>_{q,\beta_0}-\big<{\beta_F}'\big>_{q,\beta_0} = q\big<{\beta_F}^2\big>_{q,\beta_0}-{\beta_S}^2,
\end{equation}
with $\U \geq 0$ for $q \geq 1$ as expected. In order to show that $\U < 0$ can be achieved in this ensemble for $q < 1$, let us consider the density of states
$\Omega(E) = \Omega_0 E^\alpha$ with $\alpha \geq -1/2$, case which is solved in Ref.~\cite{Umpierrez2021}. In that case we have
\begin{equation}
\beta_S = \beta_0\big(1 + (\alpha+1)(1-q)\big),
\end{equation}
and
\begin{equation}
q\big<{\beta_F}^2\big>_{q,\beta_0,\alpha} = \left[\frac{(1-q)\alpha + 1}{(1-q)(\alpha+1)+1}\right]{\beta_S}^2,
\end{equation}
so
\begin{equation}
\U = {\beta_S}^2\left[\frac{q-1}{1+(1-q)(\alpha+1)}\right]
\end{equation}
which is negative for $q < 1$. For the case $q > 1$ the condition $\U \geq 0$ imposes
\begin{equation}
q \leq 1 + \frac{1}{\alpha+1},
\end{equation}
precisely is the bound reported by Lutsko and Boon~\cite{Lutsko2011}. Because $q \geq 1$ implies $\dF \leq 0$ and ${\beta_\Omega}' \leq 0$, it follows also that in this case
$\Var{\bO}{S} \geq \Var{\bF}{S}$ so this confirms that $q \geq 1$ are supercanonical-B states. On the contrary, states with $q \leq 1$ have
\begin{equation}
\label{eq:ineq_qsub}
\Var{\bF}{S} \geq \Var{\bO}{S},
\end{equation}
so they are subcanonical-A states. The inequality in \eqref{eq:ineq_qsub} follows from the results in Ref.~\cite{Umpierrez2021},
\begin{equation}
\frac{\alpha+1}{\alpha}\big<\beta_\Omega^2\big>_{q,\beta_0} = q\big<\beta_F^2\big>_{q,\beta_0}
\end{equation}
using the fact that \[q\left(\frac{\alpha}{\alpha+1}\right) < q \leq 1.\]

\subsection{The Gaussian ensemble with a convex entropy region}

Let us consider a Gaussian ensemble written as
\begin{equation}
\rho(E; \lambda) = \frac{1}{\eta(\lambda)}\exp\left(-\frac{\lambda E^2}{2}\right),
\end{equation}
where $\lambda > 0$ and with the same density of states $\Omega(E) = \Omega_0 E^\alpha$. Its fundamental inverse temperature is given by
\begin{equation}
\beta_F(E; \lambda) = -\frac{\partial}{\partial E}\ln \rho(E; \lambda) = \lambda E,
\end{equation}
and its derivative
\begin{equation}
{\beta_F}'(E; \lambda) = \lambda
\end{equation}
is non-negative, therefore it cannot be described by superstatistics. Nevertheless, we will show that it encompasses all the classes described in Table~\ref{tbl:classif}. The
$n$-th moment of the energy distribution is
\begin{equation}
\big<E^n\big>_S = \sqrt{\Big(\frac{2}{\lambda}\Big)^n}\frac{\Gamma\big(\frac{1}{2}(\alpha+n+1)\big)}{\Gamma\big(\frac{1}{2}(\alpha+1)\big)}
\end{equation}
so we can construct
\begin{equation}
\Var{\bF}{S} = \big<(\lambda E)^2\big>_S -\lambda^2\big<E\big>_S^2
= \frac{\alpha+1}{\lambda}-\frac{2}{\lambda}\left[\frac{\Gamma(\alpha/2 + 1)}{\Gamma(\alpha/2 + 1/2)}\right]^2
\end{equation}
and finally obtain
\begin{equation}
\U = \lambda\left[\alpha - \left(\frac{\Gamma(\alpha/2 + 1)}{\Gamma(\alpha/2 + 1/2)}\right)^2 \right].
\end{equation}

This is a monotonically increasing function of $\alpha$ going from $-2\lambda/\pi$ to $-\lambda/2$, thus it is always negative and all states belong
to the subcanonical class. The variance of $\bO$ is
\begin{equation}
\Var{\bO}{S} = \alpha^2 2^{\frac{\alpha-3}{2}}\lambda^{\frac{1-\alpha}{2}}\Gamma\Big(\frac{\alpha-1}{2}\Big) - \left(\frac{2}{\lambda}\right)^\alpha\Gamma(1+\alpha/2)^2
\end{equation}

In the particular case $\lambda = 2$ we can verify that there is a transition from subcanonical-A for $\alpha < 1$ to subcanonical-B behavior for $\alpha > 1$. However,
as $\alpha \rightarrow \infty$ in the thermodynamic limit we have subcanonical-B behavior for all $\lambda$.

\begin{figure}[h!]
\begin{center}
\includegraphics[width=0.55\textwidth]{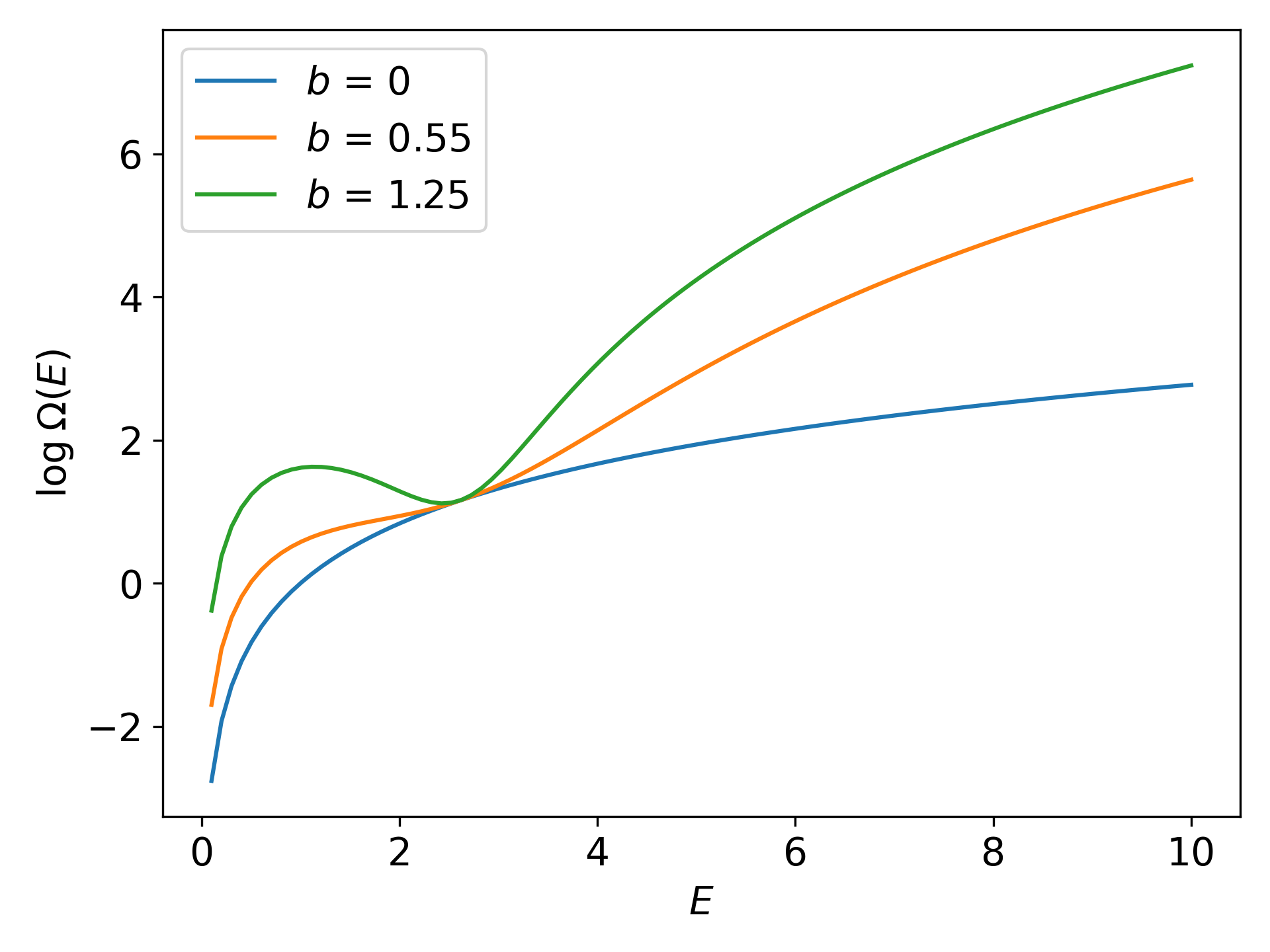}
\end{center}
\caption{Logarithm of the density of states in \eqref{eq:dos} for $\mu$=2.60384, $\alpha$=1.20401 and different values of $b$.}
\label{fig:dos}
\end{figure}

\noindent
Now let us generalize the density of states to the following expression,
\begin{equation}
\label{eq:dos}
\Omega(E; \alpha, \mu, b) = \Omega_0\big(1+b^2(E-\mu)^2\big)E^{\alpha},
\end{equation}
which is mostly concave as in the previous case but has a convex region for $b > 0$, as seen in Fig.~\ref{fig:dos}, what has been called a ``convex intruder''~\cite{Ispolatov2001,
Chavanis2002,Behringer2006,Junghans2008}. Its microcanonical inverse temperature is given by
\begin{equation}
\beta_\Omega(E) = \frac{2b^2(E-\mu)}{1+b^2(E-\mu)^2} + \frac{\alpha}{E},
\end{equation}
with derivative
\begin{equation}
{\beta_\Omega}'(E) = \frac{2b^2(b^2(E-\mu)^2-1)}{(1+b^2(E-\mu)^2)^2} - \frac{\alpha}{E^2}.
\end{equation}

\begin{figure}[h!]
\begin{center}
\includegraphics[width=0.45\textwidth]{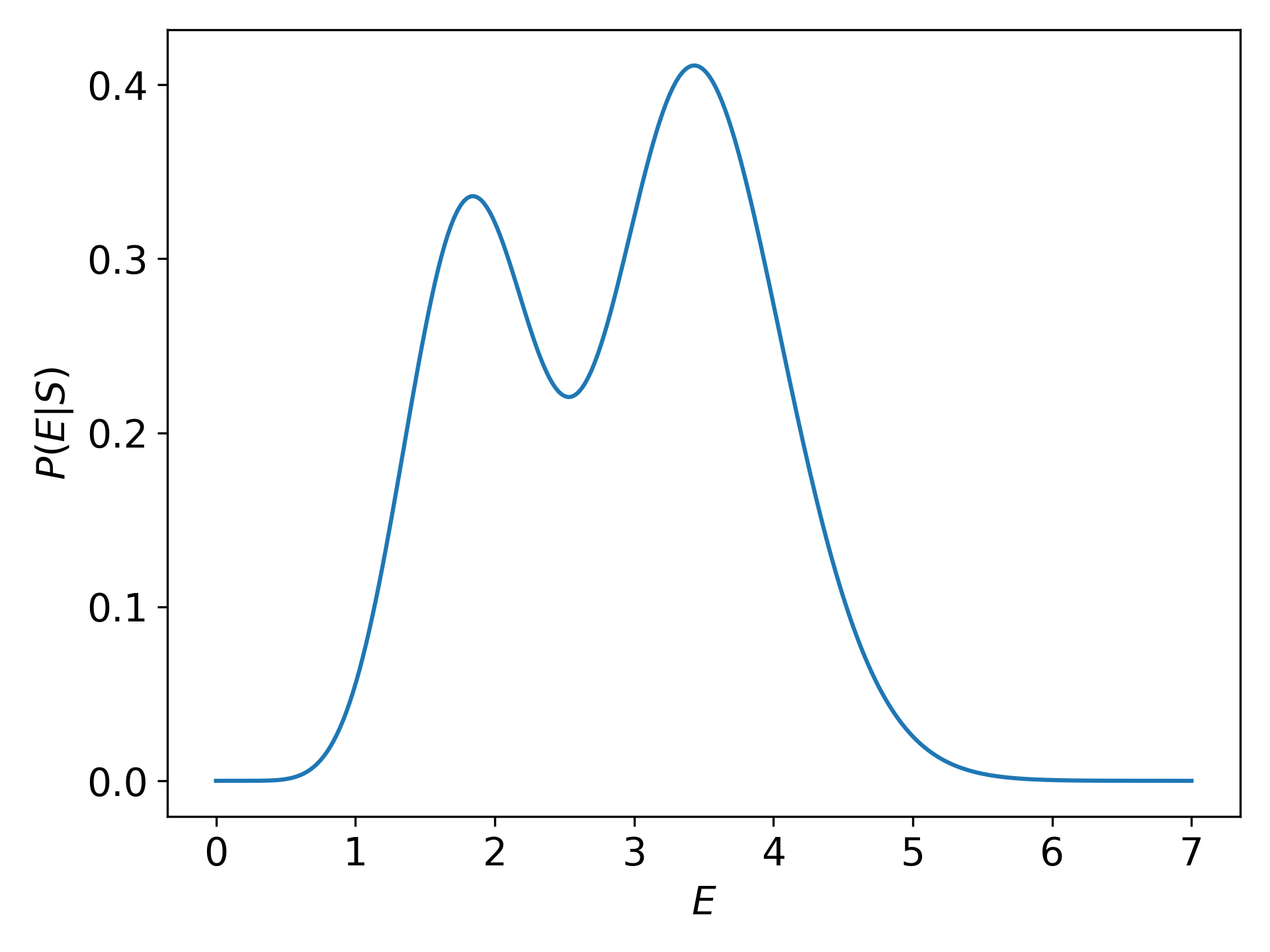}
\includegraphics[width=0.45\textwidth]{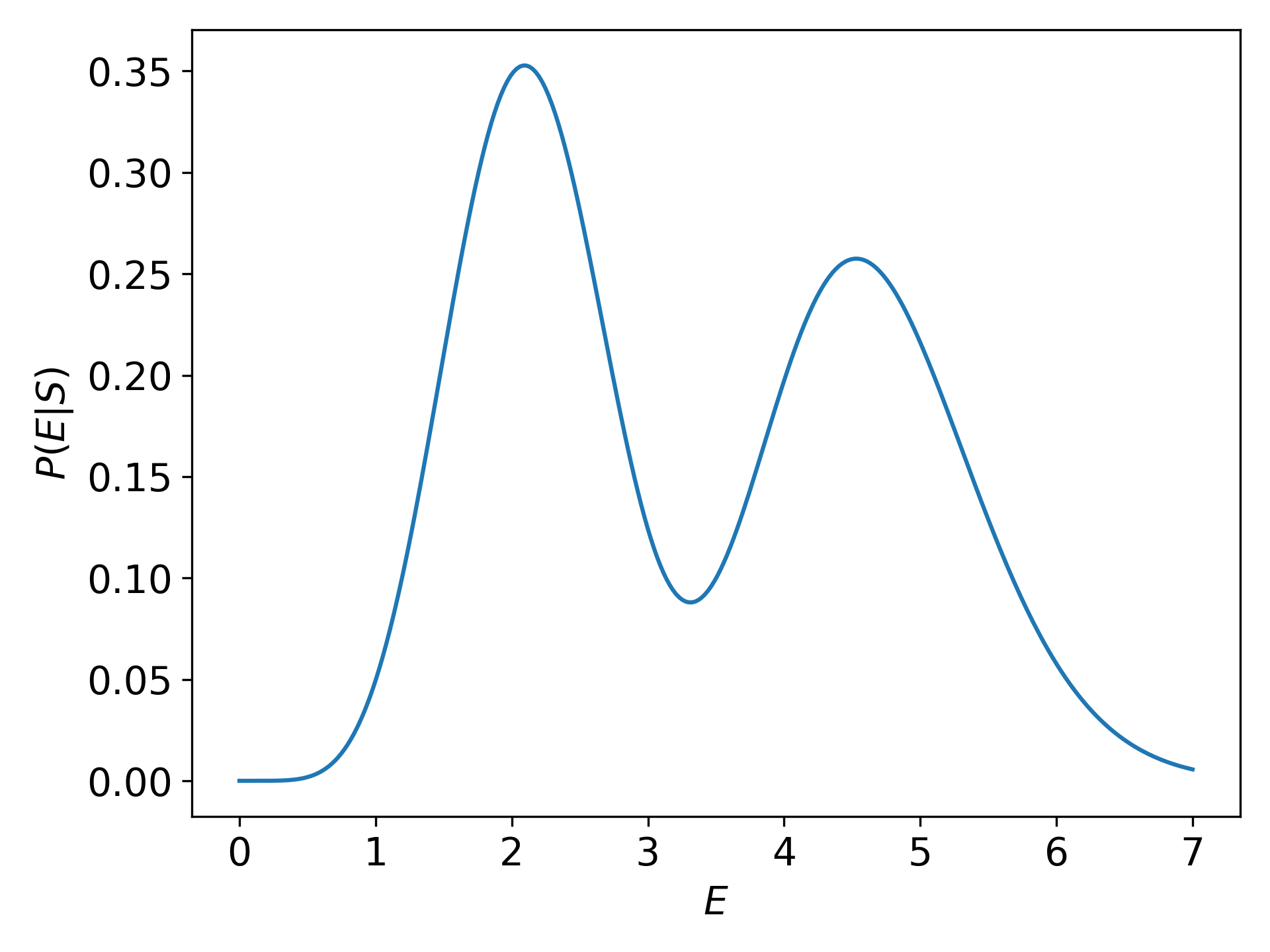}
\end{center}
\caption{Left panel, example of a supercanonical A state, with $b$=1.94541, $\lambda$=1.09747, $\mu$=2.53849 and $\alpha$=7.15724. In this case we have $\U$=0.050368. Right
panel, example of a supercanonical B state, with $b$=2.17105, $\lambda$=0.58616, $\mu$=3.28635 and $\alpha$=5.62514. In this case we have $\U$=0.13170.}
\label{fig:super}
\end{figure}

\begin{figure}
\begin{center}
\includegraphics[width=0.45\textwidth]{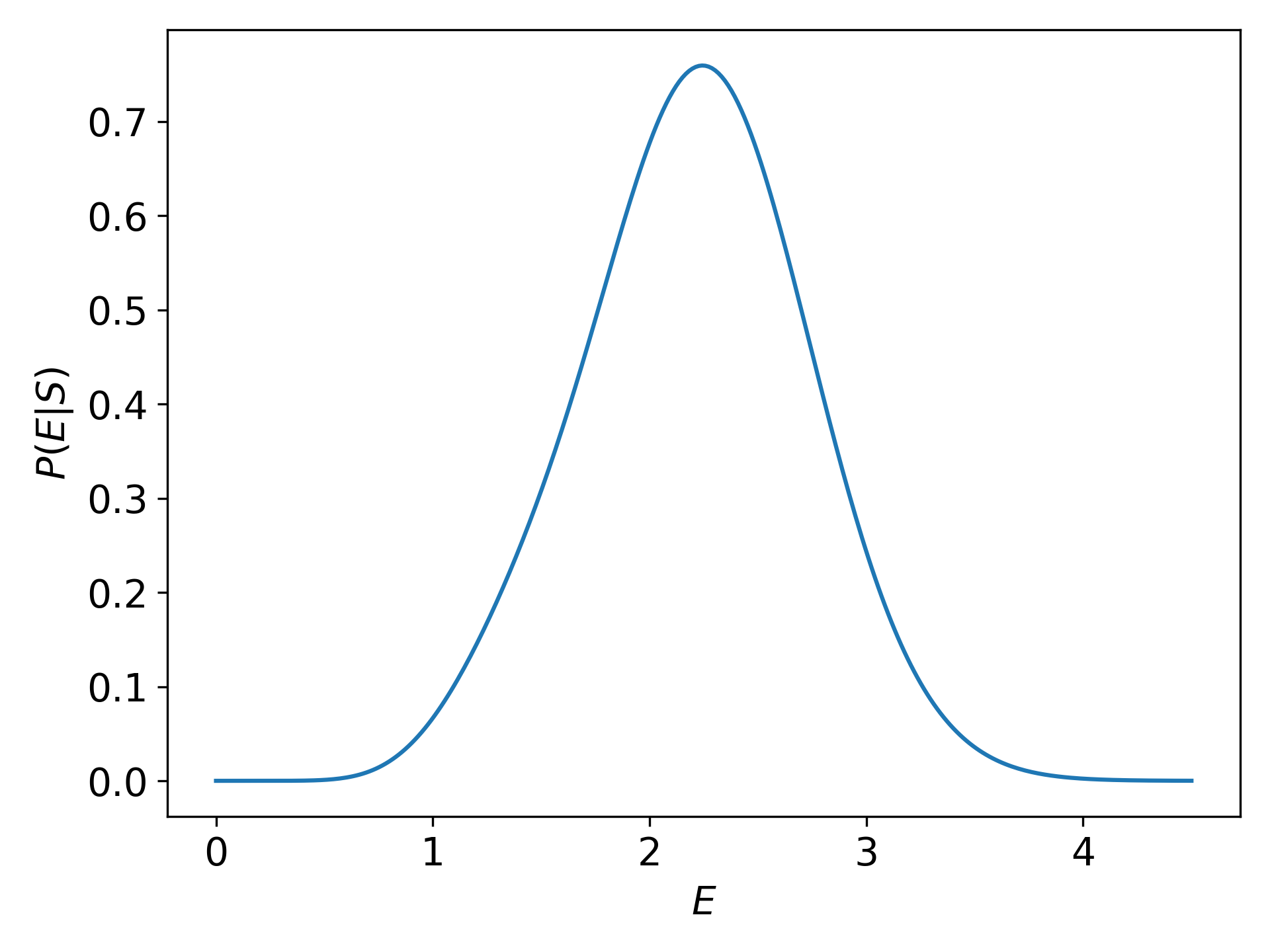}
\includegraphics[width=0.45\textwidth]{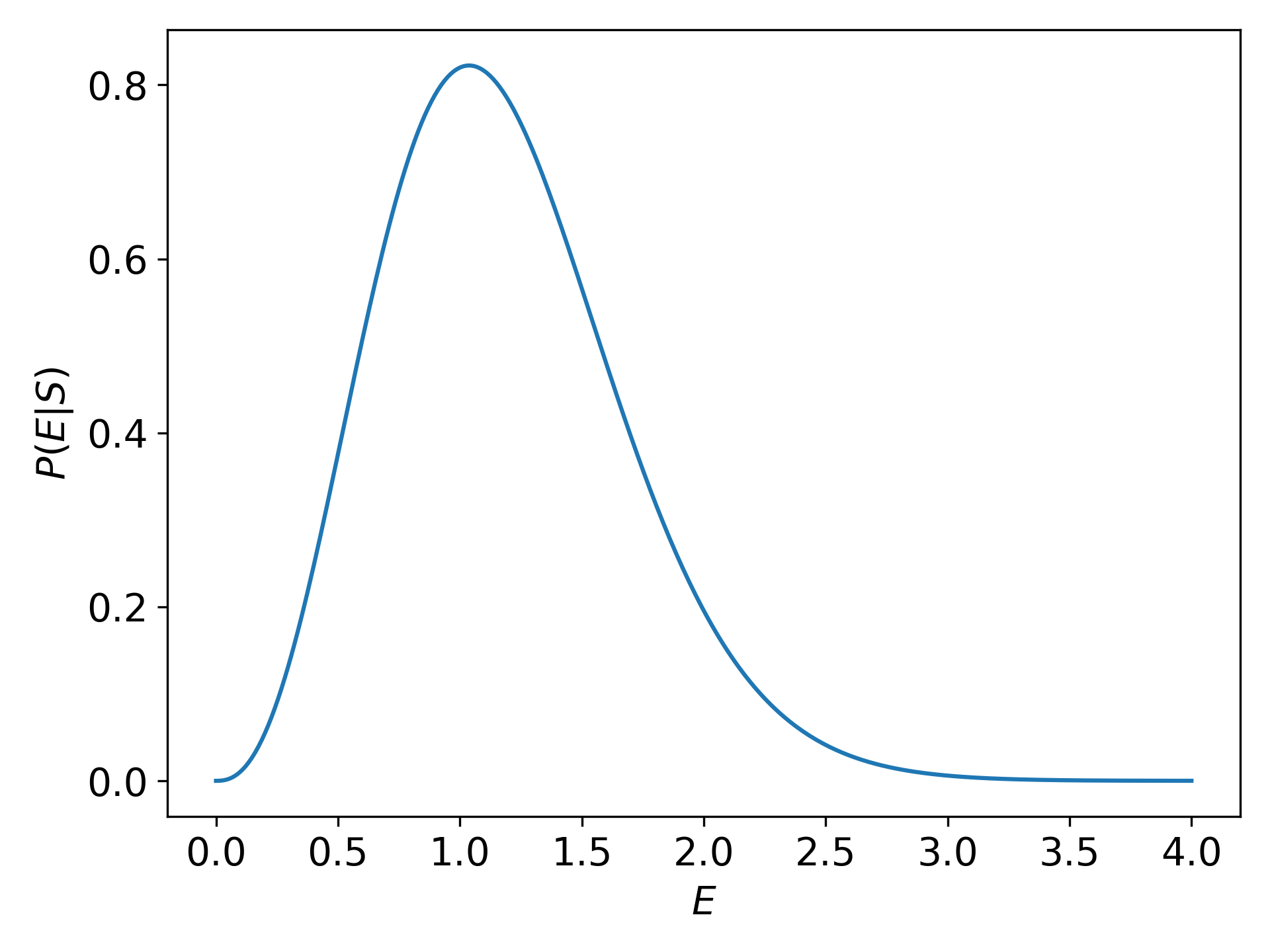}
\end{center}
\caption{Left panel, example of a subcanonical A state, with $b$=1.404639, $\lambda$=2.27263, $\mu$=1.50866 and $\alpha$=8.29722. In this case we have $\U$=-0.82574.
Right panel, example of a subcanonical B state, with $b$=0.594146, $\lambda$=1.73940, $\mu$=2.73567 and $\alpha$=2.48991. In this case we have $\U$=-1.00845.}
\label{fig:sub}
\end{figure}

\begin{figure}
\begin{center}
\includegraphics[width=0.5\textwidth]{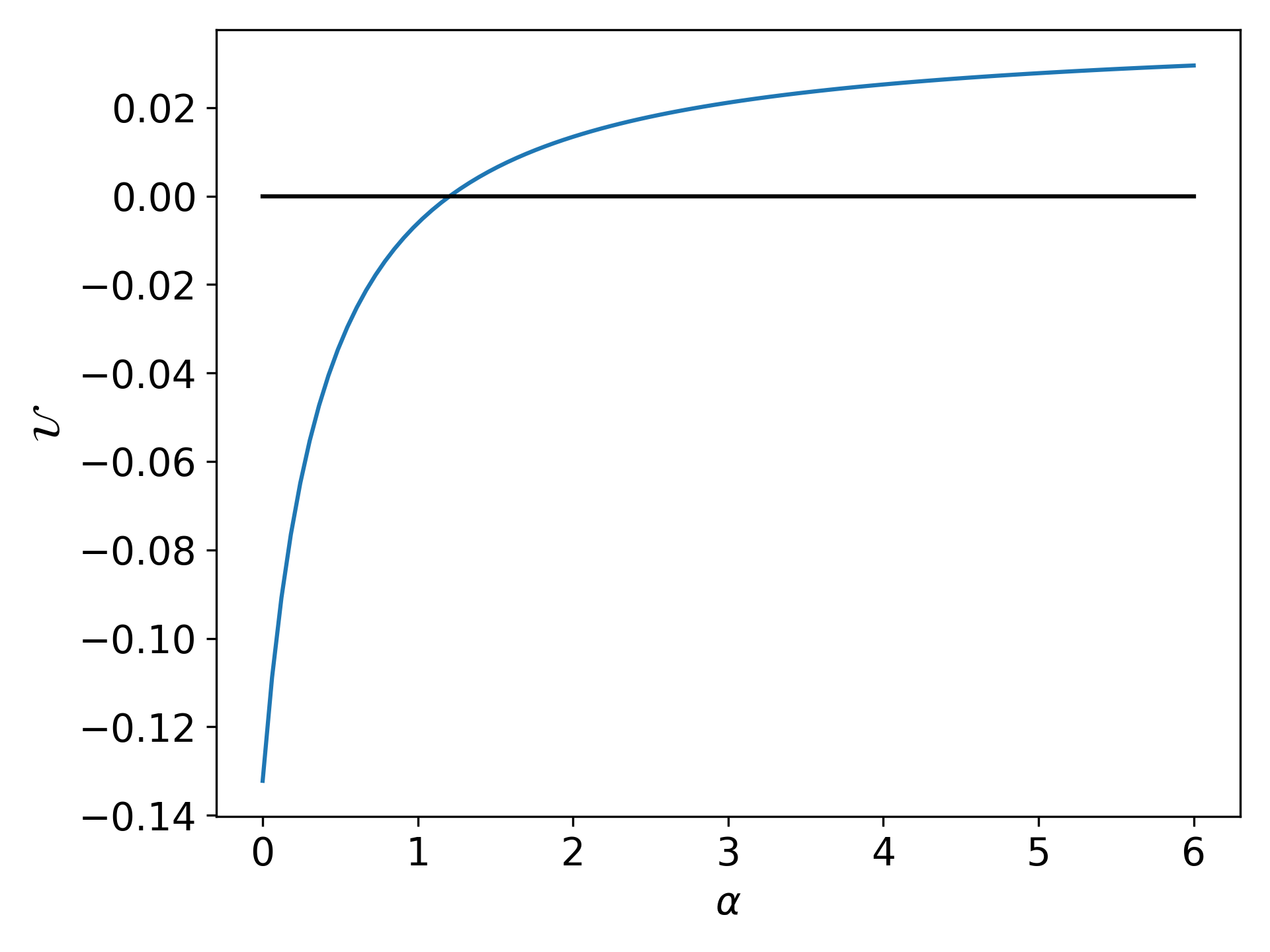}
\end{center}
\caption{Transition from subcanonical to supercanonical behavior.}
\label{fig:cross1}
\end{figure}

\begin{figure}
\begin{center}
\includegraphics[width=0.45\textwidth]{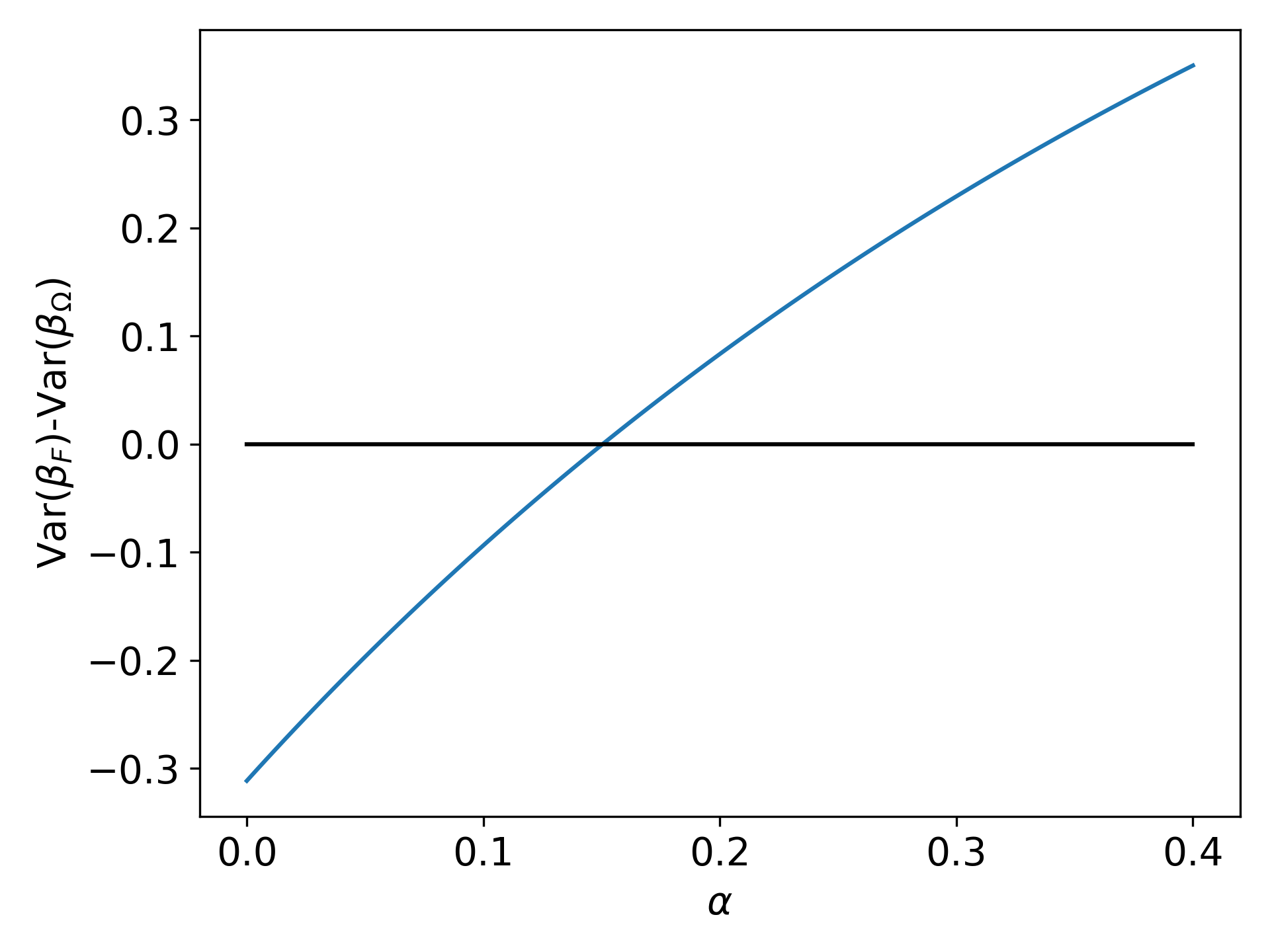}
\includegraphics[width=0.45\textwidth]{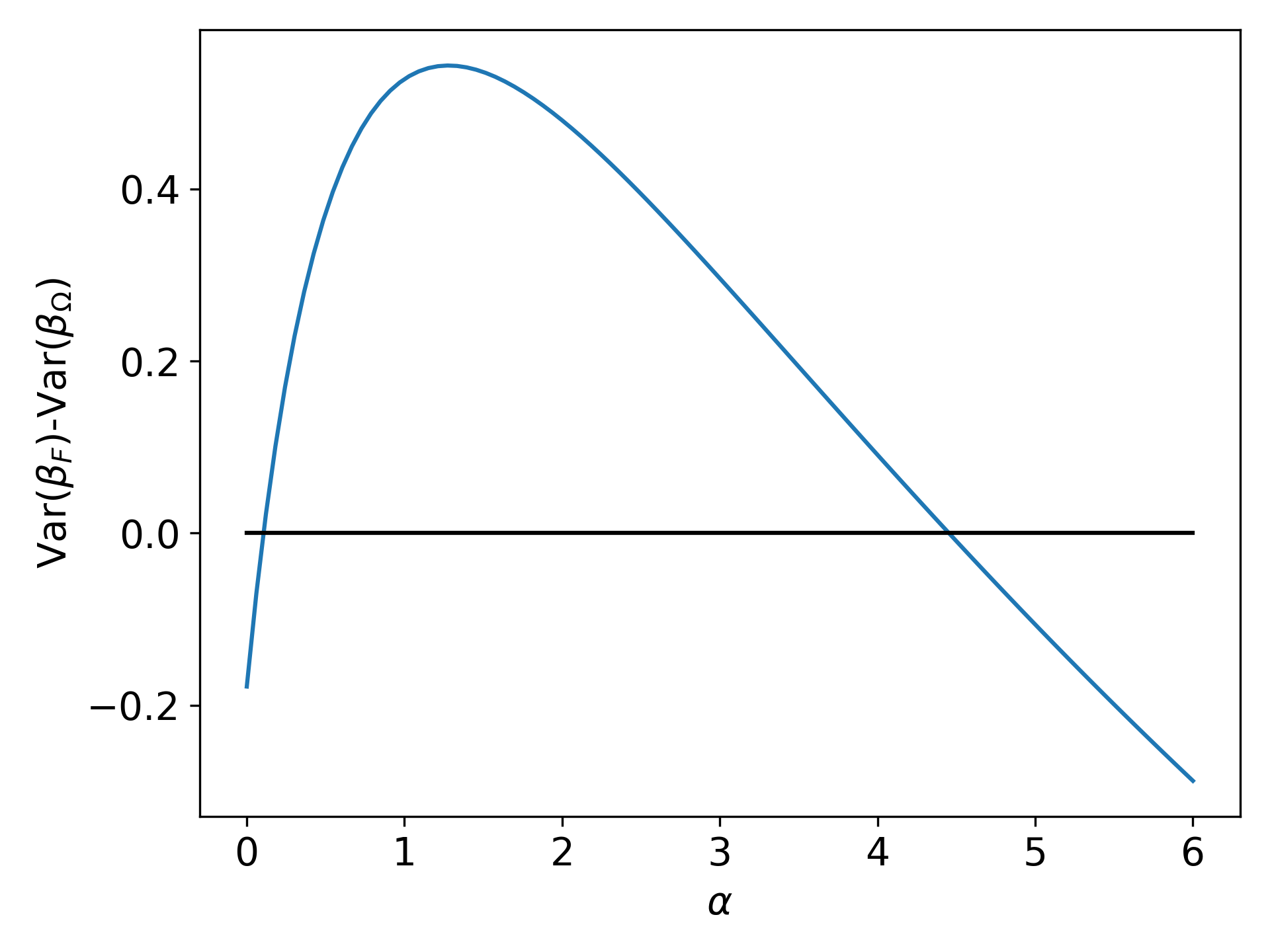}
\end{center}
\caption{Left panel, transition from subcanonical B to subcanonical A behavior. Right panel, transition from supercanonical B to supercanonical A behavior.}
\label{fig:cross2}
\end{figure}

Here we see that ${\beta_\Omega}'(E)$ can change sign for certain values of $b$, $\mu$ and $\alpha$, and thus it can describe systems with
negative and positive $C_E$. This is then a candidate for a model that can present supercanonical-A states, of which we have not shown
an example yet. In fact, such an example with a bimodal distribution of energy is seen in Fig.~\ref{fig:super} (left panel). Figs.~\ref{fig:sub},
\ref{fig:cross1} and \ref{fig:cross2} show that subcanonical states and even transitions from one class to another are also possible within this model.

\section{Concluding remarks}
\label{sec:concluding}

In this work, a classification of nonequilibrium steady states is proposed, based on the sign of the inverse temperature covariance $\U$ between inverse temperatures.
This classification divides the space of nonequilibrium models into supercanonical ($\U > 0$) and subcanonical ($\U < 0$) models, regarding all the superstatistical
models as part of the larger supercanonical-B class, disconnected from the supercanonical-A and subcanonical classes.

The main theorem proved in this work, namely the equality in \eqref{eq:equalU}, indicates that all regions of a system in a nonequilibrium steady state must
have the same value of $\U$. This is remarkably similar to the defining property of temperature in thermodynamics, the fact that $\beta$ reaches the same value
for every region of a system in thermal equilibrium, but instead connects the uncertainties $\Var{\bF}{S}$ and $\Var{\bO}{S}$ in different regions of a system.

Several results of general interest were also proven, such as the statement that all regions of a large enough homogeneous, microcanonical system with short-range
interactions under strictly monotonic $\beta_\Omega$ must have $\U \leq 0$ in a steady state, and therefore fall outside the domain of superstatistics. Moreover,
the results in this work show that supercanonical behavior is not limited to superstatistics, but can ocurr outside of it for systems with negative heat capacity,
for instance metastable states in first-order phase transitions.

\section*{Acknowledgments}

SD acknowledges financial support from ANID FONDECYT 1220651 grant.

\section*{References}

\bibliography{classif}
\bibliographystyle{unsrt}

\newpage

\end{document}